\documentclass[fleqn,usenatbib]{mnras}
\usepackage{newtxtext}
\usepackage{txfonts}
\usepackage[T1]{fontenc}

\DeclareRobustCommand{\VAN}[3]{#2}
\let\VANthebibliography\thebibliography
\def\thebibliography{\DeclareRobustCommand{\VAN}[3]{##3}\VANthebibliography}

\usepackage{graphicx}	
\usepackage{url}
\usepackage{courier}
\usepackage{helvet}

\title[]{Investigation on upstream ion events from L1 point observation: New Insights}

\author[B. Dalal et al.]{
Bijoy Dalal,$^{1}$\thanks{E-mail: bijoydalal.at@gmail.com}
Dibyendu Chakrabarty,$^{1}$
Christina M. S. Cohen,$^{2}$
and Nandita Srivastava$^{3}$
\\
$^{1}$Physical Research Laboratory, Ahmedabad - 380009, India\\
$^{2}$California Institute of Technology, Pasadena, CA 91125, USA\\
$^{3}$Udaipur Solar Observatory, Physical Research Laboratory, 
Udaipur - 313001, India
}

\date{Accepted XXX. Received YYY; in original form ZZZ}

\pubyear{2024}

\begin{document}
\label{firstpage}
\pagerange{\pageref{firstpage}--\pageref{lastpage}}
\maketitle

\begin{abstract}
Origin of energetic upstream ions propagating towards the Sun from the Earth’s bow shock is not understood clearly. In this letter, relationship between solar wind suprathermal and upstream ions has been investigated by analyzing fluxes of H, $^4$He, and CNO obtained from multidirectional in-situ measurements at the first Lagrange point of the Sun-Earth system during 2012--2014. 49 upstream events have been selected based on flux enhancements of the upstream ions in comparison with the solar wind suprathermal ions. An energy cut-off at $<$ 300 keV is observed for the upstream events. This is attributed to the efficacy of the particle acceleration process near the bow shock. Interestingly, spectra of upstream ions soften systematically as compared to the spectra of their solar wind counterpart with decreasing mass of elements. The degree of spectral softening increases with decreasing mass-to-charge ratio of the species. Since during most of the events the interplanetary magnetic field was radial, we argue that cross-field diffusion of upstream ions gives rise to the modulation (spectral softening) of upstream ions, which is dependent on the mass-to-charge ratio of species. Our work indicates towards a systematic change in solar wind suprathermal ions after interaction with the bow shock.     
\end{abstract}




\section{Introduction}

Origin and characteristics of energetic ions ($<$ 1 MeV) upstream of the Earth’s bow-shock has been studied for the past few decades (e.g., \citealp{Asbridge_et_al_1968, Sarris_et_al_1976, Scholer_et_al_1979, Paschmann_et_al_1980, Desai_et_al_2000, Meziane_et_al_2002, Kronberg_et_al_2011} etc.). These upstream events are generally characterized by short durations ($\sim$1-2 hours), steeply falling spectra (j$\sim$ E$^{-2}$  $^{to -6}$ \citep{Anagnostopoulos_et_al_1998}, field aligned sunward anisotropies (\citealp{Mitchell_and_Roelof_1983, Muller_et_al_2008, Desai_et_al_2008} etc.). These events are observed in higher numbers associated with corotating interaction regions (CIRs) (e.g., \citealp{Desai_et_al_2000}), intense geomagnetic activity \citep{Anagnostopoulos_et_al_1998}. Despite a number of studies, there exists significant uncertainty about the origin of these upstream ions. There are broadly two schools of thought regarding the origin of these upstream particles--(i) these ions are accelerated by the Earth’s bow shock \citep{Lee_1982, Trattner_et_al_2003} and (ii) these are generated in the Earth’s magnetosphere \citep{Sheldon_et_al_2003, Anagnostopoulos_et_al_2005, Chen_et_al_2005} and diffuse upstream of the bow shock. 

There are different types of ions present in the upstream region of the Earth’s bow shock. Specularly reflected ions are the least energetic (with energies of few keV) among all the suprathermal and energetic ions that are observed immediate upstream and downstream of the quasi-perpendicular (shock normal angle, $\theta _Bn > 45^o$) bow shock. These ions are energized by the solar wind electric field \citep{Gosling_et_al_1982} and are confined to within one gyro-radius upstream from the shock \citep{Gosling_and_Robson_1985}. A relatively more energetic ($< \sim 10$ keV) field aligned proton beams are thought to be produced either by reflection from the shock or by leakage of hot proton populations from the magnetosheath into the upstream region \citep{Schwartz_and_Burgess_1984}. Diffuse ions, on the other hand, can extend up to 200 keV per charge and show almost isotropic distribution in the spacecraft frame \citep{Scholer_et_al_1979, Scholer_et_al_1981}. Large-amplitude hydromagnetic waves are always observed in association with these diffuse ions \citep{Paschmann_et_al_1979}.       

One of the interesting observations regarding upstream ion events is that these ions are observed  at wide spatial distances starting from  the vicinity of the bow-shock \citep{Ipavich_et_al_1981, Meziane_et_al_2002} to up to $1750R_E$ \citep{Desai_et_al_2008, Kronberg_et_al_2011} upstream of it. \cite{Paschmann_et_al_1980} proposed a theory in which solar wind particles are reflected by the Earth’s bow shock and the energy gained by these particles is a function of interplanetary magnetic field, solar wind velocity, and the local shock normal. Interestingly, \cite{Dwyer_et_al_2000} reported that there is more than 50$\%$ chance that two spacecraft located at $\sim$ $70R_E$ apart can also observe upstream events simultaneously. These authors remarked that the size of the source regions of these simultaneously observed upstream events are larger than the spatial size of the bow shock. While most of these studies indicated a magnetic connection between the spacecraft and the bow shock to be essential for observing upstream events, \cite{Haggerty_et_al_2000} has reported upstream events that were observed with transverse magnetic field configuration ahead of the bow shock. Results by \cite{Desai_et_al_2008} show that upstream events can be observed simultaneously by two spacecraft even if they are separated by $800 R_E$ both radially and laterally. These authors suggested that one of the possible sources of these events is large amplitude Alfv\'en waves with spatial scales of the order of 0.03 AU.  

Upstream ion distributions in the vicinity of the bow shock and far upstream are very different \citep{Scholer_et_al_1981}. It is possible that highly anisotropic energetic particles far-upstream of the bow shock are originally the leaked diffuse ions that travelled mostly scatter free to long distances \citep{Scholer_et_al_1981, Mitchell_and_Roelof_1983}. However, the particle fluxes decay exponentially with increasing distance from the bow shock \citep{Ipavich_et_al_1981, Trattner_et_al_1994, Kis_et_al_2004}. These studies essentially show that the e-folding distance in that case depends on the energy of the particles indicating energy dependent escape of particles from the bow shock \citep{Desai_et_al_2008}. How the properties of ions with different masses are modulated during the transport from the bow shock to interplanetary (IP) medium remains unclear till date. 

One way to address this issue is to explore  any systematic relationship that may or may not exist between the solar wind suprathermal  ions (propagating away from the Sun) and the upstream ions (propagating towards the Sun) reaching the L1 orbit. Since there are evidences of the solar wind ion populations to be scattered by the Earth’s bow shock, upstream ions should carry the signatures of bow shock modulation. Therefore, a comparative study between suprathermal particles during upstream events at L1 and solar wind suprathermal particles crossed L1 earlier might capture these effects. We conjecture that this approach may lead to two scenarios about upstream ions. One possible scenario may be systematic modulation (changes in the spectral slope of upstream ions in different events very systematically) indicating dominance of bow shock related effects and the second scenario may be random modulations of spectral slopes indicating multiple processes/sources in action that includes bow shock as well.  Keeping this objective in mind, by using the observations from the Wind spacecraft, we compare spectra of suprathermal H, $^4$He, and CNO propagating towards the Sun and towards the Earth. The analyses bring out a mass-dependent systematic softening in the spectra of sunward propagating suprathermal ions over the suprathermal ions propagating toward the Earth.

\section{Data used and instrumentation}
H, $^4$He, and CNO fluxes analyzed in this work are obtained from the two telescopes (T1 and T2) of the Suprathermal Energetic Particle (STEP) instrument of the Energetic Particles: Anisotropy, Composition, and Transport (EPACT) investigation \citep{von_Rosenvinge_et_al_1995} on board the Wind spacecraft. The STEP instrument is capable of measuring $^4$He--Fe in the energy range of 0.03--2.0 MeV per nucleon and H in the 0.1--2.5 MeV energy range \citep{von_Rosenvinge_et_al_1995, Desai_et_al_2000}. T1 and T2 of the STEP instrument are oriented at $\pm$26$^o$, respectively with respect to the ecliptic plane (equivalently, 116$^o$ and 64$^o$, respectively with respect to the spacecraft’s spin axis, which is along –Z direction in the Geocentric Solar Ecliptic (GSE) coordinate system. T1 and T2 scan the interplanetary medium just above and below the ecliptic plane in eight azimuthal sectors of 45$^o$ each (see \citealp{Desai_et_al_2000} for a schematic of the scan planes and sector information). This configuration allows us to analyze and compare suprathermal particles that come from both the ``Sun sector” (i.e., Sun-looking sector) (by combining data of sector 1, 2, and 3 of STEP telescopes) and the ``bow shock sector'' (bow-shock-looking sector) (by combining data from sector 5, 6, and 7 of the STEP telescopes). In the following section, we discuss the selection criteria and characteristics of upstream events analyzed in this work.

\begin{table}
	\centering
	\caption{List of events with start time, end time, specie concerned, and the enhancement ratio (defined in the text)}
	\label{tab1}
	\begin{tabular}{lcccc} 
		\hline
		No. & Start Time  & End Time & Specie & $\frac{F_{BS}}{F_{SS}} \times 100$ \\
		\hline
		1 &	2012-01-06 06:59 &	2012-01-06 07:53 &	H   & 393.749\\
        2 &	2012-01-16 16:59 &	2012-01-16 18:03 &	H   & 1512.27\\
        3 &	2012-01-23 06:21 &	2012-01-23 07:59 &	$^4$He & 114.834\\
        4 &	2012-02-02 10:17 &	2012-02-02 11:32 &	$^4$He	& 84.4822\\
        5 &	2012-02-17 07:11 &	2012-02-17 08:48 &	H	& 779.042\\
        6 &	2012-04-27 09:43 &	2012-04-27 10:26 &	$^4$He	& 549.542\\
        7 &	2012-05-20 00:29 &	2012-05-20 01:34 &	$^4$He	& 63.926\\
        8 &	2012-05-23 14:06 &	2012-05-23 16:27 &	H	& 226.302\\
        9 &	2012-07-07 09:56 &	2012-07-07 11:16 &	$^4$He	& 538.512\\
        10 & 2012-08-01 01:34 &	2012-08-01 01:57 &	$^4$He	& 394.995\\
        11 & 2012-08-20 07:06 &	2012-08-20 08:00 &	H	& 696.153\\
        12 & 2012-08-21 04:48 &	2012-08-21 06:05 &	$^4$He	& 315.105\\
        13 & 2012-08-27 04:39 &	2012-08-27 05:12 &	H	& 1019.99\\
        14 & 2012-09-05 17:03 &	2012-09-05 17:56 &	H	& 157.573\\
        15 & 2012-09-20 16:57 &	2012-09-20 17:50 &	$^4$He	& 379.765\\
        16 & 2012-10-11 12:12 &	2012-10-11 13:16 &	$^4$He	& 366.06\\
        17 & 2012-10-14 04:16 &	2012-10-14 04:59 &	H	& 757.895\\
        18 & 2012-11-26 20:42 &	2012-11-26 21:58 &	$^4$He	& 158.763\\
        19 & 2013-01-08 02:34 &	2013-01-08 04:03 &	$^4$He	& 333.546\\
        20 & 2013-01-27 13:58 &	2013-01-27 15:23 &	$^4$He	& 160.476\\
        21 & 2013-01-27 20:48 &	2013-01-27 22:14 &	$^4$He	& 794.423\\
        22 & 2013-03-13 06:25 &	2013-03-13 07:08 &	H	& 156.743\\
        23 & 2013-03-28 00:44 &	2013-03-28 01:37 &	H	& 993.808\\
        24 & 2013-05-05 05:56 &	2013-05-05 06:47 &	H	& 188.689\\
        25 & 2013-06-03 23:32 &	2013-06-04 01:20 &	$^4$He	& 372.638\\
        26 & 2013-06-04 10:45 &	2013-06-04 12:34 &	H	& 375.331\\
        27 & 2013-06-21 02:06 &	2013-06-21 03:11 &	$^4$He	& 315.827\\
        28 & 2013-06-21 04:17 &	2013-06-21 04:59 &	$^4$He	& 384.996\\
        29 & 2013-06-30 08:54 &	2013-06-30 09:38 &	$^4$He	& 85.5203\\
        30 & 2013-07-18 23:23 &	2013-07-19 00:27 &	$^4$He	& 328.526\\
        31 & 2013-08-22 09:55 &	2013-08-22 11:21 &	$^4$He	& 166.405\\
        32 & 2013-10-16 21:20 &	2013-10-16 22:46 &	H	& 149.947\\
        33 & 2013-11-10 10:54 &	2013-11-10 12:52 &	$^4$He	& 334.968\\
        34 & 2013-11-10 17:56 &	2013-11-10 19:22 &	$^4$He	& 321.981\\
        35 & 2013-11-17 18:22 &	2013-11-17 19:37 &	$^4$He	& 1156.59\\
        36 & 2013-12-08 21:47 &	2013-12-09 01:46 &	$^4$He	& 155.697\\
        37 & 2013-12-31 14:51 &	2013-12-31 15:25 &	$^4$He	& 290.382\\
        38 & 2014-01-03 11:46 &	2014-01-03 12:40 &	$^4$He	& 143.064\\
        39 & 2014-01-04 01:34 &	2014-01-04 03:00 &	$^4$He	& 492.851\\
        40 & 2014-01-04 03:13 &	2014-01-04 03:57 &	H	& 654.446\\
        41 & 2014-03-06 15:27 &	2014-03-06 16:21 &	$^4$He	& 606.834\\
        42 & 2014-04-04 12:46 &	2014-04-04 13:42 &	$^4$He	& 394.748\\
        43 & 2014-05-04 15:26 &	2014-05-04 16:20 &	$^4$He	& 485.102\\
        44 & 2014-06-19 15:13 &	2014-06-19 18:16 &	$^4$He	& 752.106\\
        45 & 2014-08-04 14:11 &	2014-08-04 14:44 &	H	& 378.574\\
        46 & 2014-10-20 19:54 &	2014-10-20 20:49 &	$^4$He	& 235.134\\
        47 & 2014-12-08 04:37 &	2014-12-08 05:40 &	$^4$He	& 171.128\\
        48 & 2014-12-08 20:47 &	2014-12-08 22:58 &	$^4$He	& 206.907\\
        49 & 2014-12-09 02:12 &	2014-12-09 03:41 &	$^4$He	& 790.912\\
		\hline
	\end{tabular}
\end{table}

\section{Selection of upstream events and method}
We select 49 upstream events based on enhancements in the H (at 0.14 MeV [in T2]), $^4$He (at 0.06 MeV per nucleon [T2]) observed in the “bow shock sector” during 2012--2014. We identify an event based on the enhancement in at least one of these two elements in the “bow shock sector” when compared to the “Sun sector” at the same time. The start and end times of these events are selected based on visual inspections. The start and end times of the events are enlisted respectively in column 2 and 3 of Table \ref{tab1}. The concerned element (H or $^4$He) used for event identification is mentioned in column 4 of Table \ref{tab1}. In all the events, we see a clear rise and fall in ion fluxes in the bow shock sector. We calculate the enhancement percentages (defined as [average flux recorded in the bow shock sector during an upstream event (F$_{BS}$)/average flux recorded in the Sun sector during the event (F$_{SS}$)] $\times$ 100, all fluxes measured by T2) for each of these events and tabulate them in column 5 of Table \ref{tab1}. The lower limit of the enhancement percentage is set to be 150$\%$ of H flux and 60$\%$ $^4$He flux observed in the Sun sector to identify an upstream event. The enhancement percentages vary in the range 156.74--1512.27 and 63.93--1156.59 for H and $^4$He, respectively, for all 49 events. It is observed that ion fluxes measured by T2 are higher than fluxes measured by T1 for most of the times. We have given preference to the flux variations observed by T2 while choosing these events. It is observed that not all the three elements exhibit enhancements during all the events mentioned here. 15 and 34 events among the 49 events listed in Table \ref{tab1} are chosen based on increment in H and $^4$He, respectively. Enhancements in multiple elements are also observed in many events.

\begin{figure}
	\includegraphics[width=\columnwidth]{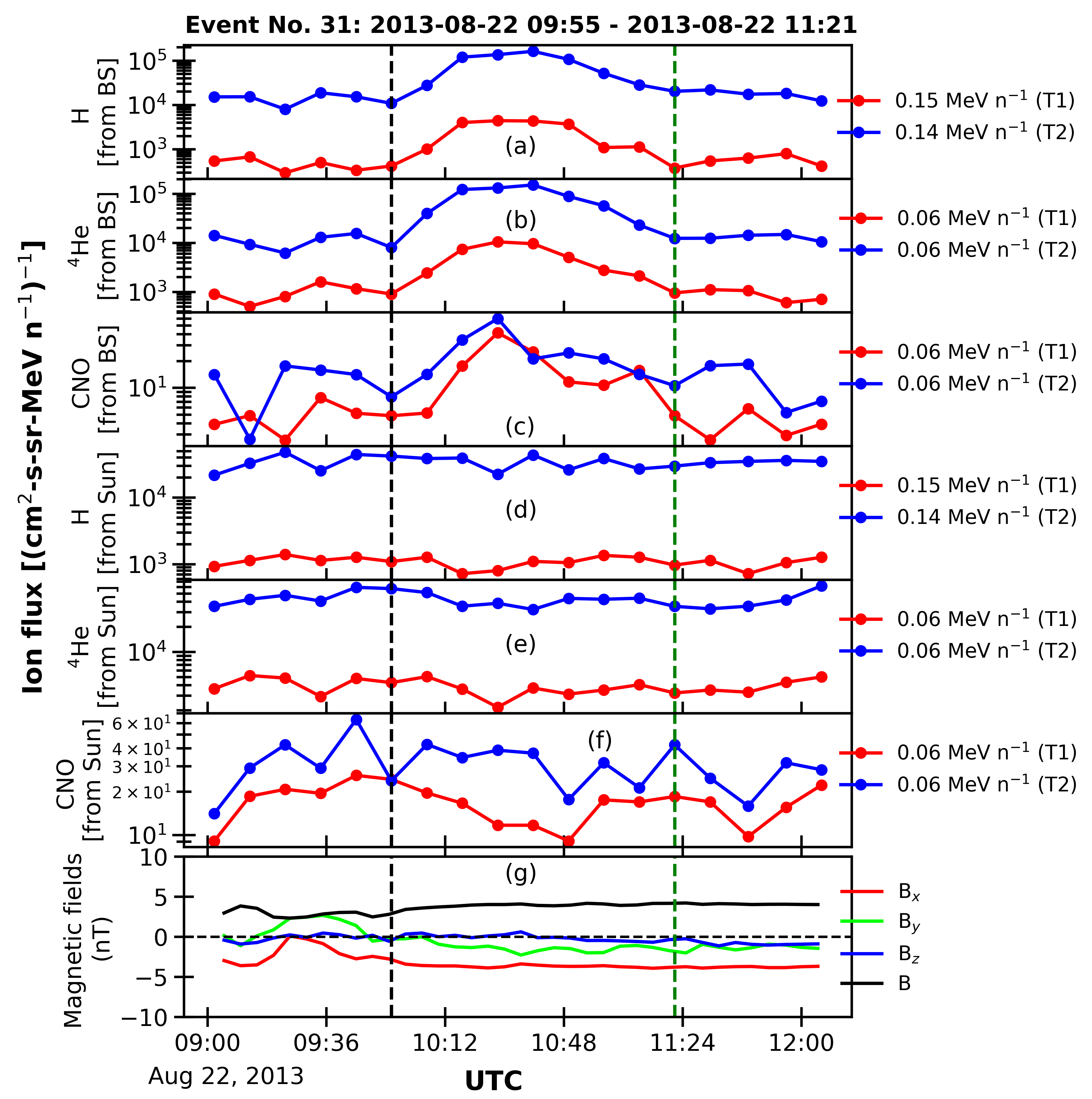}
    \caption{Flux variations during the upstream event 31 as listed in Table \ref{tab1} Panel (a), (b), and (c) show respectively H, CNO, and $^4$He fluxes measured by T1 (blue) and T2 (red) from the bow shock direction. The energy values are mentioned in the right side of each panel. H, CNO, and $^4$He fluxes coming from the “Sun sector” are shown in panel (d), (e), and (f), respectively. Interestingly, T2 reports greater fluxes than that measured by T1 in most of the events. The start and end times of the event are marked by vertical black and green dashed line, respectively. As mentioned in Table \ref{tab1}, this event is chosen based on variation in $^4$He (at 0.06 MeV per nucleon) flux measured by T2. Panel (g) shows the components of interplanetary magnetic field (red: $B_x$, green: $B_y$, and blue: $B_z$) and total magnetic field, $B$ (black) in the GSE coordinate system.}
    \label{fig1}
\end{figure}

Figure \ref{fig1} captures a typical event in which temporal variations of H, $^4$He, and CNO fluxes from the “bow shock sector” and the “Sun sector” corresponding to upstream event 31 in Table \ref{tab1} are shown. The top three panels (a, b, and c) clearly show enhancements in H, $^4$He, and CNO fluxes during the event (interval between the black and green vertical dashed lines) above the background. The next three panels (e, f, and g) show flux variations in the “Sun sector” of the telescopes. The enhancement percentage for $^4$He is calculated to be 166.41. Components of IMF in geocentric solar ecliptic (GSE) coordinate system during the event are shown in panel (g) of Figure \ref{fig1}. It is seen that the IMF connecting the spacecraft and the nose of the bow shock is nearly radial and lies in the ecliptic plane during the event. Once the upstream events are selected, we calculate the spectral indices of all the three elements if there is enhancement in at least three consecutive energy channels. The results obtained from this analysis are discussed in the following section. 

\begin{figure}
	\includegraphics[width=\columnwidth]{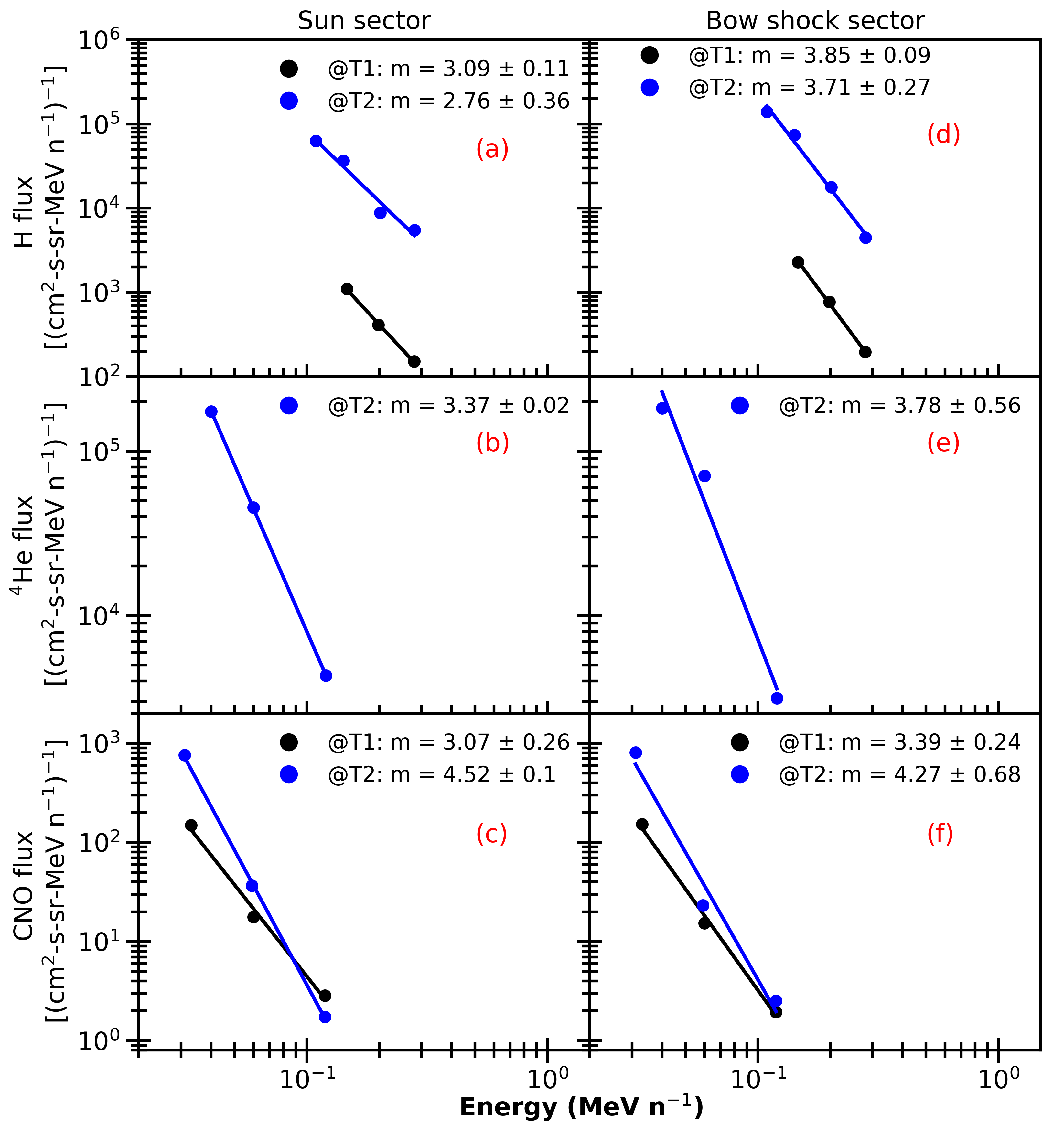}
    \caption{Spectra of H, $^4$He, and CNO during upstream event No. 31 observed by T1 (black) and T2 (blue) in the “Sun sector” are shown in panel (a), (b), and (c), respectively. The corresponding spectra for the “bow shock sector” are shown in panel (d), (e), and (f). The spectral indices and standard errors (SEs) are mentioned in the upper right corner of the panels.}
    \label{fig2}
\end{figure}

\section{Results}
\begin{figure}
	\includegraphics[width=\columnwidth]{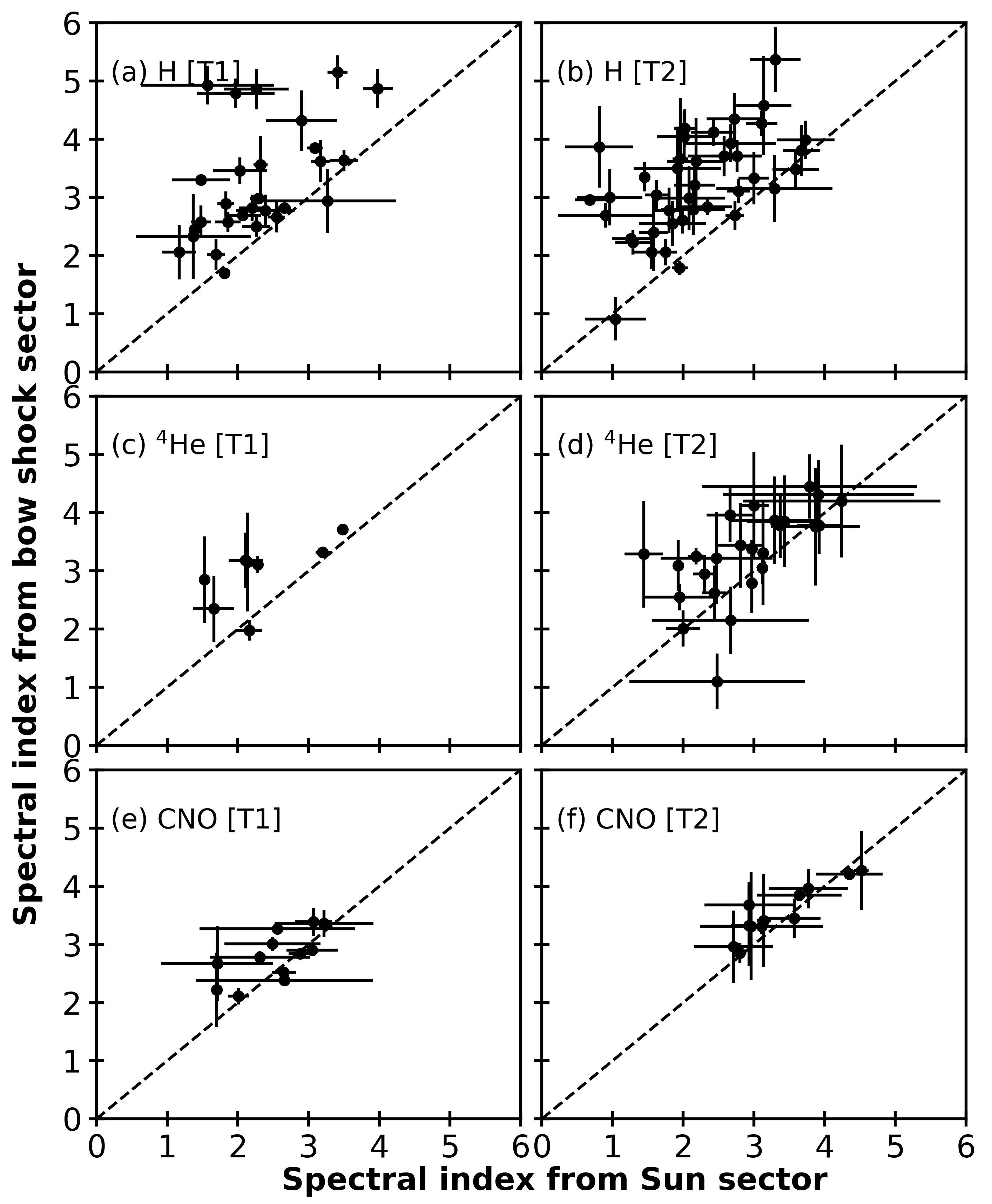}
    \caption{Comparisons of spectral indices of H [(a) T1, (b) T2]; $^4$He [(c) T1, (d) T2], and CNO [(e) T1, (f) T2] observed from the “Sun sector” and “BS sector” of the telescopes. The black $45^o$ dashed line in each panel represents the line of equal spectral indices.}
    \label{fig3}
\end{figure}

In this section, we present a comparison between the spectral indices of H, $^4$He, and CNO observed in the “Sun sector” and the “bow shock sector” during the selected upstream events. In order to do so, we assume that the Sun is the major source of particles and the “bow shock sector” of T1 and T2 observes the particles scattered/reflected by the bow-shock. We acknowledge the fact that there is possibility of mixing of particles observed at L1 from different sources. Considering the path of a particle (say, H) to be a straight line between the position of the Wind satellite and the bow-shock nose with a path length of 200 $R_E$, we can calculate the travel-time of a proton with kinetic energy E from the spacecraft location to the bow shock nose and back. For E= 60 keV, this travel-time appears to be around 25 minutes. Therefore, according to our assumption, particles detected in the “Sun sector” of the detector units at any point of time are expected to be detected by the “bow shock sector” of the detector units on an average 30 minutes later. We show in Figure \ref{fig2} the spectra of H, $^4$He, and CNO observed at the “Sun sector” 30 minutes before the upstream event 31 and “bow shock sector” during the event by T1 and T2 telescopes. Ion fluxes are averaged over the same durations as those of upstream events while calculating the spectral indices from the “Sun sector”. From the spectral indices mentioned in Figure \ref{fig2} we find that spectra observed in the ‘bow shock sector” seem to be softer than that observed in the “Sun sector” of T1 and T2. This aspect is further illustrated in Figure \ref{fig3} where we plot spectral indices of H, $^4$He, and CNO observed at the “bow shock sector” against those observed at the “Sun sector” of T1 and T2 separately. The $45^o$ black dashed line in each panel of Figure \ref{fig3} indicates equal spectral indices in both the sectors. From panel (a) and (b) of Figure \ref{fig3} we can see that spectral indices of H in the “bow shock sector” are most of the times above the line of equal spectral indices (systematic softening) in T1 and T2, respectively. Interestingly, spectral indices of CNO (panel ‘e’ and ‘f’) observed in both T1 and T2 in the “bow shock sector” go almost hand in hand (no softening) with the spectral indices observed in the “Sun sector”. It can be seen from panel (d) of Figure \ref{fig3} that the softening in $^4$He spectra observed in the “bow sector” looks intermediate between H (softest among the three elements) and CNO (least soft). This mass-dependent hardening effect is more prominent in case of T2.  

\begin{figure}
	\includegraphics[width=\columnwidth]{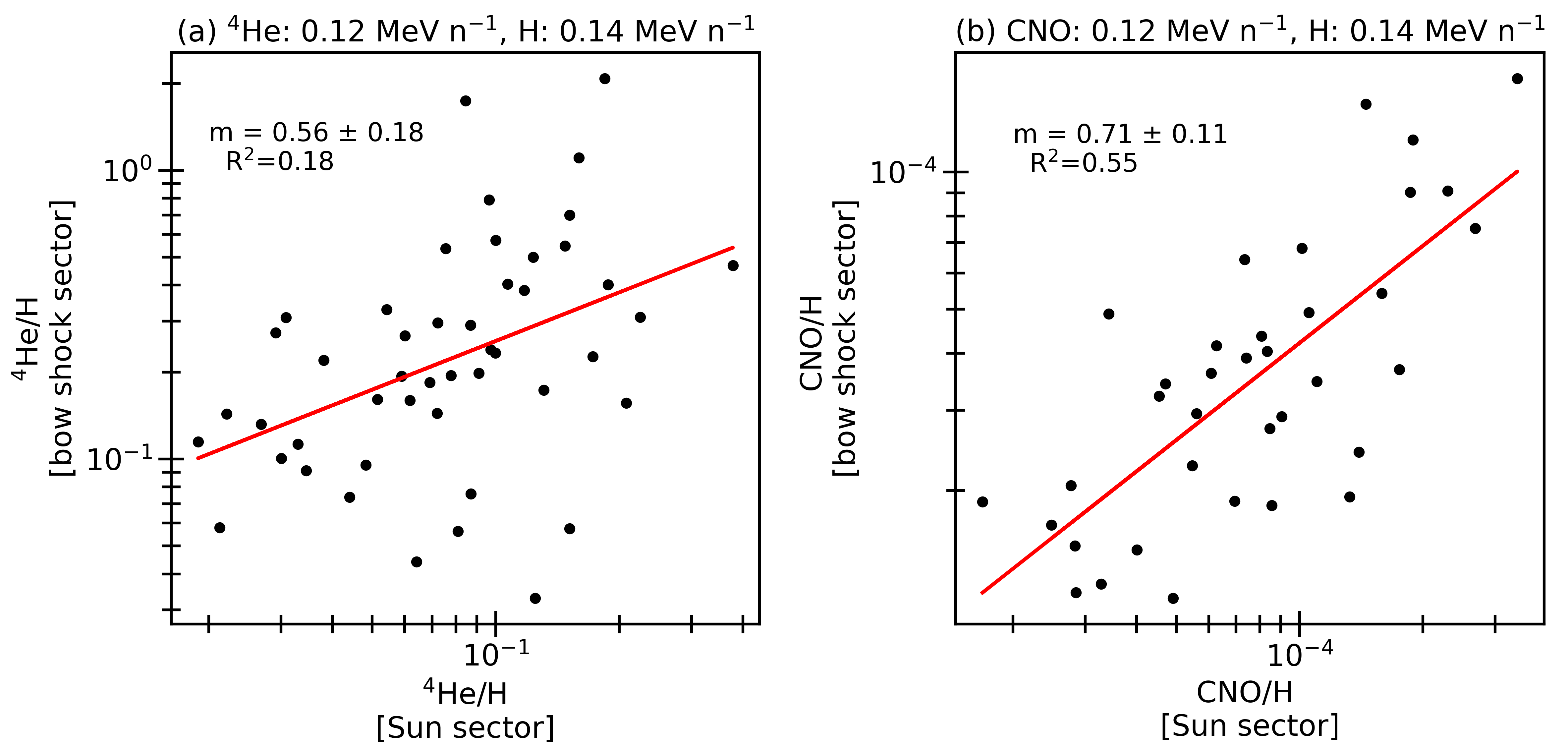}
    \caption{(a) Comparisons between the $^4$He/H ratios observed at the Sun sector 30 minutes earlier than the start of upstream events (durations being same as that of the upstream events) and $^4$He/H ratios during the upstream events. All fluxes are measured by T2 telescope. The ratio is calculated for nearly equal energy channels of $^4$He and H. The energy channels are mentioned at the top of the panel. The red line shows the linear regression fit. The corresponding fit parameters (slope, $m$ and goodness of fit, $R^2$) are mentioned in the panel. (b) Same as panel (a), but for CNO/H. }
    \label{fig4}
\end{figure}
  
In order to evaluate the origin of the upstream ions ($^4$He and CNO), we compare $^4$He/H and CNO/H between the Sun sector and the bow shock sector in Figure \ref{fig4}. This figure shows that the correlation between (CNO/H)Sun sector and (CNO/H)bow shock sector (see panel ‘b’, $R^2$=0.55) is better than the corresponding correlation between ($^4$He/H)Sun sector and ($^4$He/H)bow shock sector (see panel ‘a’, $R^2$=0.18). The implications of these observations are discussed in the following section.

\section{Discussions and conclusions}
The durations of these upstream events discussed in the previous section vary in the range of 0.5--3.5 hours, which is consistent with the existing literature (e.g., \citealp{Desai_et_al_2000}. The degree of enhancements in H, $^4$He, and CNO fluxes differ and vary from event to event, which is clear from Table \ref{tab1}. CNO fluxes show enhancements in fewer events as compared to H or $^4$He fluxes. This may be due to lack of solar wind CNO fluxes above the instrument threshold level before many of those events occur.

In the present work, upstream events exhibit steeply falling spectra with spectral indices in the range 2--6. This is consistent with earlier results (\citealp{Anagnostopoulos_et_al_1998, Desai_et_al_2000} etc.). Further, it is seen that in 80$\%$ of the upstream events observed by T2, when spectral indices of H have been calculated, the enhancement in H fluxes is limited to $<$ 300 keV per nucleon. The corresponding percentages for $<$ 300 keV per nucleon cut-off for CNO and $^4$He are 100$\%$ and 93$\%$, respectively. Therefore, it appears that there is an energy cut-off of the upstream events observed in the “bow shock sector”. \cite{Meziane_et_al_2002} showed by statistical study of upstream events that in the absence of preexisting population of energetic ions ($>$ 50 keV), ion energy spectrum is limited to 200-300 keV. It is verified that at least 42 among 49 upstream events analyzed in this study neither do follow an interplanetary coronal mass ejections (ICMEs) nor associated with high-speed solar wind stream, which are considered as sources of energetic particles in the solar wind. In such cases, \cite{Meziane_et_al_2002} did not find any dependence of the energy spectrum on shock geometry. Turbulent quasi-parallel bow shock can accelerate particles up to $\sim$ 300 keV almost isotropically \citep{Lin_et_al_1974, Greenstadt_et_al_1980}. \cite{Ellison_and_Mobius_1987} showed that the fraction of solar wind energy flux imparted to the high-energy ion flux (the injected particles) falls off very rapidly with energy. This limits the ion acceleration up to very high energy. Diffuse ions are the most energetic among all types of upstream ions discussed in the introduction section. Therefore, an energy cut-off at 300 keV per nucleon suggests that the observed upstream events are essentially diffuse ions accelerated by the first order Fermi acceleration at the quasi-parallel bow shock and propagated to the orbit of the Wind spacecraft.  
It can be seen from Figure \ref{fig3} that the spectral indices of H observed in the “bow shock sector” of both T1 and T2 are more scattered and above the line of equal spectral indices. This suggests a systematic softening of the H spectra observed in the “bow shock sector”. This is a new way to look at the upstream ions. The difference between spectral indices for both the sectors decreases with increase in the mass of the species (H $>$ $^4$He $>$ CNO). Figure \ref{fig3} also reveals that spectral softening decreases with increasing mass of the elements. This mass-dependent modulation in spectra of energetic particles observed upstream of the bow shock at L1 point is also new. \cite{Sarris_et_al_1978} hinted towards rigidity-dependent escape of protons from the magnetosphere and subsequent acceleration of these particles to explain harder proton spectra outside the magnetosphere than spectra observed inside the magnetosphere. In this study, we see a mass-dependent modulation in particle spectra during upstream events at L1 point. Since the observed upstream ions appear to be the diffuse ions, the proposition of Sarris et al. does not seem to explain our observations. Instead, we feel that the softening of particle spectra is due to mass-dependent modulation of the ions during the course of their transport from the bow shock to the spacecraft. One of the reasons for such modulation could be cross-field diffusion of these upstream ions. An estimate of the cross-field diffusion due to resonant scattering shows \citep{Jokipii_1987}

\begin{equation}
    {\frac{k_{\perp}}{k_{\parallel}}} = \frac{1}{1+\left(\frac{\lambda_{\parallel}}{r_{g}}\right)^2}
\end{equation}

where $k_{\parallel}$ and $k_{\perp}$ are the coefficients for diffusion parallel and perpendicular to the magnetic field, $\lambda_{\parallel}$ is the parallel mean free path and $r_{g}$ is ion gyro-radius, which depends on the mass-to-charge ratio ($m/q$) of ion species. $r_{g}$ increases with increasing $m/q$, so does $\frac{k_{\perp}}{k_{\parallel}}$ resulting in greater mixing of upstream ions with the solar wind suprathermal ions. Therefore, the points mostly lie along the line of equal spectral indices between the Sun sector and bow shock sector for CNO.  On the contrary, in case of H, the cross-field diffusion is less, which allows the upstream ions reach the L1 point in a focused manner.  Since bow shock softens the solar wind spectra, the points mostly lie above the line of equal spectral indices. The behavior of $^4$He seems to be intermediate between H and CNO. This suggests that cross-field diffusion may play an important role during the propagation of upstream ions from bow shock to the L1 point. Cross-field diffusion also supports observation of upstream events in places far apart laterally consistent with earlier observations (\citealp{Dwyer_et_al_2000, Desai_et_al_2008} etc.). Not only in upstream events, \cite{Dalal_et_al_2022} have shown the modulation of quiet-time suprathermal particles depending on $m/q$ of elements. Therefore, it appears that whenever modulation of suprathermal particles in the solar wind magnetic field is concerned, it is dependent on the $m/q$ of the species.   

Figure \ref{fig4} seems to give an idea about the sources of the upstream ions. The good correlation between CNO/H from the “Sun sector” and the “bow shock sector” probably suggests that upstream CNO reaching L1 are mostly of solar wind origin. On the contrary, poor correlation between $^4$He from the “Sun sector” and the “bow shock sector” suggests that $^4$He fluxes reaching at L1 get mixed up with suprathermal populations coming from other sources. Among these sources may be $^4$He ions leaked from the magnetosphere, pick-up $^4$He ions etc. Detailed investigations are needed to understand this aspect.  

It is observed that IMF was near radial (see panel ‘g’ of Figure \ref{fig1} for an example) during 67$\%$ of the events analyzed here. This is consistent with earlier studies (e.g., \citealp{Desai_et_al_2000}). Radial magnetic field is indicative of connection between the L1 point and the magnetosphere of the Earth \citep{Haggerty_et_al_2000}. It is possible that these upstream ions propagated along the magnetic field in all the events. This proposition appears valid because the typical width of magnetic flux tubes linked to the Earth in case of impulsive solar energetic particle (ISEP) events is $\sim$ $4.7 \times 10^6$ km ($\sim$ 750 $_RE$) \citep{Mazur_et_al_2000, Giacalone_et_al_2000}. The spatial scale length in which interplanetary shocks accelerate particles is $\sim$ $2.97 \times 10^6$ km ($\sim$ 460 $R_E$) \citep{Neugebauer_et_al_2006}. Therefore, large-scale magnetic flux tubes seem to be very common in the heliosphere. Our study essentially supports the observation of upstream events originated from a small source region and thereafter propagating along magnetic flux tubes with larger spatial scales.  

Although we observed systematic softening of spectra as far as the upstream ions are concerned with respect to solar wind, we cannot rule out the possibility of mixing of suprathermal particles from multiple sources like diffused ions from the magnetosphere, bow shock accelerated particles, particles accelerated by reverse shocks of stream interaction regions (SIRs) formed beyond the Earth’s orbit, pick up ions etc. In this work, we have not explored this aspect. In near future, we intend to shed more light on this aspect using multidirectional data obtained from the Aditya Solar wind Particle Experiment (ASPEX) (e.g., \citealp{Goyal_et_al_2018}) payload on board the Aditya-L1 spacecraft. 

\section*{Acknowledgements}
We are grateful to the Principal Investigator (PI) and all the members of the STEP team for constructing the STEP instrument and thereafter, generating and managing the dataset used in this work. We express our sincere gratitude to the Department of Space, Government of India for supporting this work. 
\section*{Data Availability}
The STEP flux data are available at \url{https://cdaweb.gsfc.nasa.gov/index.html}. In addition to the STEP flux data, we have used data for the components of interplanetary magnetic field (IMF) that are available in the above-mentioned site.


\bibliographystyle{mnras}
\bibliography{Paper_3_BD} 

\bsp	
\label{lastpage}
\end{document}